# Design of Backscatter Tailored Optical Fibers for distributed magnetic field sensing using Fiber Optic Pulsed Polarimetry

Roger J. Smith
*1121 Lupine, Lake Forest, California 92630*

A fiber optic pulsed polarimeter utilizes a LIDAR return signal from locations within a single mode optical fiber for *local* magnetic field measurements. Discrete fiber Bragg gratings are written along the fiber to produce a sequence of robust reflections from a short polarized light pulse propagating down the fiber. The fiber can be placed in free space or any region tolerating its introduction. This *backscatter tailored optical fiber*(BTOF) produces, up to physical and practical constraints, a signal encoding the progressive Faraday rotation along the fiber once placed in a $\boldsymbol{B}$ field, inversion of which yields the *local* component of $\boldsymbol{B}$, $B_{\|}(s)$ to a design spatial resolution and accuracy. The reflection sequence is abstract and adapts to any length of fiber by assigning a spacing between FBGs. Multipathing can contaminate the prompt LIDAR signal, and the effect of 3rd order *non-local* back-reflections for a uniformly spaced but arbitrary reflection sequence is calculated, allowing BTOF designs to meet accuracy requirements with confidence. The performance of the BTOF can be validated experimentally. Design criteria are given for uniform reflection and flat return BTOF designs. Mathematical algorithms are derived, and the *Narayana* numbers are found to be involved.

## I. INTRODUCTION

Magnetic Fusion Energy devices (tokamak, FRC, RFP, stellarator, gas dynamic traps, magnetized target fusion) require a knowledge of the boundary magnetic field adjacent to the plasma volume. High temporal BW is important for *real–time* feedback and control and high spatial resolution is needed for characterizing the edge $B$ field dynamics: high poloidal mode numbers and field perturbations due to the discreteness of the magnetic coils, support structures producing eddy currents in the vacuum shell and the plasma's response. Advanced tokamak operations may use *non*-axisymmetric coils to produce field perturbations(RMP) at the plasma edge for ELM control(1,2) and the plasma's response needs to be well characterized. Electrical sensor solutions are limited by constraints that can be overcome with fiber sensors.

The measurement choice is between using electrical probes (pickup and Rogowski coils) based on magnetic induction or fiber optic magnetic sensors based on the Faraday effect. Fiber optic sensors have upfront costs for a laser and polarimeter but have several distinct advantages and are recognized on the ITER project as viable. The advantages to using optical fibers are: the magnetic field, $\boldsymbol{B}(t)$ or current, $I(t)$, strengths are directly sensed over a bandwidth from DC to GHz, with no saturation effects and high fidelity, electrical probes measure $\mathrm{d}\boldsymbol{B}/\mathrm{d}t(=B_{\mathrm{dot}})$ and $\mathrm{d}I/\mathrm{d}t$ requiring electronic conditioning integrators, filters and multiplexers, introducing processing artifacts that require careful characterization to remove; electrical high voltage hazards, noise due to electromagnetic interference, electrostatic coupling and ground loops are avoided; the diagnostic footprint is small and non-perturbative, fibers can multiplex many optical signals on one fiber return while electrical probe arrays require bulky wire bundles that lead to EM coupling between individual sensors. Large conduits dedicated to distributed electrical probes become a small diameter tube allowing generous placement around the device. Disadvantages of optical fiber sensors include: a temperature dependent Verdet constant; possible signal perturbation from polarimetric artifacts due to fiber movements (vibrations); and ellipticity due to residual stress in forming the fiber and from bending the fiber. A fiber optic pulsed polarimeter (FOPP) provides the highest spatial density of field measurements. Potentially, 100-200 or more local field measurements can be obtained with a *space–time* field resolutions of '10*cm–ns*'. It is conceivable to cover the boundary of the device so thoroughly that the *Poynting* flux, the rate of energy supplied to the plasma volume from the external circuits, can be accurately determined, complementing the plasma's stored energy from internal measurements.

The FOPP measurement concept applies to a plethora of high magnetic field and EM pulsed devices, such as rail guns, plasma guns, plasma shocks, plasma accelerator, science investigating intense transient magnetic fields and recently, the development of high temperature superconducting magnets[3]. Plasma dynamics: magnetic reconnection and dynamical *non*-linear plasma behavior can be *non*-perturbatively investigated to very fine scales in *time* and *space*. FOPP can resolve vacuum $B$ fields to 1% of *ambient field@space–time* resolutions: 10T@*mm*–10*ps*, 100G@10*cm*–*ns* and 10G@*m*–0.1$\mu$s. The size of the sensor allows *distributed* measurements of $B(s)$ along the fiber within a volume allowing unprecedented intimacy to the field source often at very high rep rates.

This paper is organized as follows: Fiber Optic Pulsed Polarimetry and BTOFs are reviewed in **Section II**; in **Section III** the 3rd order reflected and 2nd, 4th order transmitted energies are calculated for Flat and Uniform BTOFs; instrumental design considerations are detailed in **Section IV**; **Section V** gives practical applications of FOPP using BTOF to MFE devices and a conclusion is given in **Section VI**. **Appendix A** details an algorithm for the number of multi-paths with odd number of reflections exiting the fiber and serendipitously arrives at the *Narayana triangular* numbers. **Appendix B** details algorithms used in this paper. MKS units are used and a guide to the notation and terminology is given in **Appendix C**.

## II Fiber optic pulsed polarimetry and BTOF

Fiber optic pulsed polarimetry[4] is a LIDAR-*like* technique used to resolve the SOP of backscatter induced by a polarized light pulse as it propagates in the fiber to measure the *distributed*, *local* magnetic field along the fiber. The FOPP was inspired by Pulsed Polarimetry[5] and later found a commonality with POTDR[6], a Raleigh backscatter fiber optic magnetic sensing technique. Rayleigh backscatter-based LIDAR is not viable for laboratory sized devices as the return power is too weak for *cm*-*m* fiber lengths, the *SNR* is too low for single shot detection. The coupling efficiency, $S$ of the pulse into backscatter is low for dipole scattering ~0.5%, most of the energy is lost to the cladding. The return power increases with $E_o$ as for Thomson scattering but in a SMF, $E_o$<1–10$\mu$J, where *non*-linear SBS or SRS takes over before fiber breakdown. The author embarked on a few failed attempts to artificially enhance the backscatter in fibers before successfully testing the first BTOF based on wavelength resonant FBGs. FBGs act as weak plane mirrors with $r+t\sim1$. FBGs can increase the back–reflection many thousandfold above Rayleigh[7]. The use of FBGs with prescribed $\lambda$–tuned reflections allow tailored measurements of $B_{||}(s)$ along the fiber to a designed spatial and magnetic field resolution for a given $\lambda$, a *backscatter-tailored optical fiber*. Robust reflections, however, can generate multiple scattering contaminating the LIDAR signal. Design criteria are presented to avoid this.

The Faraday effect relates to the accumulated change, $\alpha(s)$ of the polarization azimuth, $\psi$ in the return signal to the *local* field component parallel to the fiber's axis, $B_{||}(s)(=\boldsymbol{B}(s)\cdot\hat{\boldsymbol{s}})$. If $\alpha_f(s)$ is the line integrated change in $\psi$ up to $s$ then $\alpha(s)$ at the polarimeter is given by Eq (1) where $s$ is given by *time-of-flight*, $s=ct/2N_r$, ($N_r$~1.5)

$$\alpha(s) = 2V_\lambda \int_0^s \boldsymbol{B}(s')\cdot d\boldsymbol{s}' \tag{1}$$

$V_\lambda$ is the Verdet constant of the fiber. The rate of rotation, $\Delta\alpha(s)/\Delta s$ is then directly related to $B_{||}(s)$, given by Eq (2), the inversion of Eq (1),

$$B_{||}(s) = \frac{1}{2V_\lambda}\Delta\alpha(s)/\Delta s \tag{2}$$

The two analyzed intensities, $I_{s,p}(t)$ are used to determine $\alpha(t)=atan\sqrt{I_p/I_s(t)}$, unwrapped. The field, $B_{||}(s)$ is not directly measured but given by the difference of two sequential α measurements, $\alpha(s)$ and $\alpha(s+\Delta s)$. If the field distribution is *quasi*-static for $\Delta t_f > 2N_rL_f/c$, fiber length $L_f$, then the backscatter SOP, $\alpha(L_f)$ is $2\alpha_f(L_f)$, given that the Faraday effect is *non*-reciprocal under reflection. For highly dynamic $B$ fields changing on a $\tau_B$ time scale, the usable length of fiber is $l_B$=(10*cm*/*ns*)$\tau_B$.

This does not preclude the FOPP being used as a FOCS to measure the forward line integrated Faraday rotation, $\alpha_T(t_p)=\alpha_f(L_f,t_p)$, directly related to $I_p(t_p)$.

Figures of merit for the fiber sensor are: 1) its transparency at $\lambda_o$ which determines a maximum useful length, $L_f$, 2) its Verdet constant, $V_\lambda$, 1 *rad*/T-*m* for silica and 20 *rad*/T-*m* for Tb-doped glass @1064*nm* with a $1/\lambda^2$ dependence and 3) maximum pulse energy which determines a limiting *SNR,* see **Section IV**.

Both fiber types are transparent at 532*nm* with the Tb-doped fiber limited to ~2*m* and silica fiber to 100's of meters. The incremental Faraday rotation, $\Delta\alpha_{si,1064nm}$=5.5°·$\Delta s_5$@1T and $\Delta\alpha_{Tb,1064nm}$=11°·$\Delta s_5$@0.1T. The sensitivity is quadrupled at $\lambda_o$=532*nm*. These two fiber types cover 1-10*m* size MFE devices well.

The SMF is used to support the FBGs, typically 20-50$\mu m$ in length, written at positions, $s_m=(m-1)\Delta s$ along the fiber with reflection coefficients, $r_m$, $t_m=(1-r_m)$ at $\lambda_o$. A distributed backward propagating signal(a timed sequence of reflections) is generated by the light pulse propagating down the fiber. The pulse length, $l_p=c\tau_p/N_r<\Delta s$, is less than the FBG separation so that reflections are isolated from each other. For $\Delta s$=5$cm$, $\tau_p$<0.25$ns$. $N_p$ pulses are produced over a duration $\Delta t_f$ =2$L_f N_r/c$, with $L_f$=($s_{Np}$–$s_1$)=($N_p$–1)$\Delta s$, with receiver, $\tau_{det}$ integrating and smoothing the modulations. The signal power, $P^1$~$r_1 E_o N_p/\Delta t_f$ scales as 1/$L_f$. Given a scaling factor, γ between two BTOFs with $\Delta s'$=γ$\Delta s$, the backscattered power increases to $P^1$/γ as $\Delta t_f'$ is compressed by γ<l. Powers from $m$W to kW are easily generated for the same pulse energy.

The fiber is the supporting structure for reflectors and is also a magneto-optically active medium. Together, the fiber supported reflectors and the fiber, itself, provide a *synergistically* robust *distributed* magnetic field sensor providing useful spatial and field resolutions spanning all fusion research from MFE to MHEDLP devices.

The backscatter is shunted to the polarimeter using a non-polarizing SMF 3 dB splitter as shown in **Fig. 1**. The BTOF can be Tb-doped SMF fused(connected) to silica SMF to provide spatial discrimination given the 20 fold difference in Verdet constants. Otherwise, the field outside the area of interest contributes to the SOP for an all–silica BTOF. After the splitter, PMF is used.

The FBG is written on the fiber using an interference pattern from two coherent UV pulses split from a parent pulse and spatially focused on the fiber core. Recently, ultrafast (sub-$ps$) NIR pulsed lasers have been used as, at this wavelength, the acrylic coating is transparent and can be left intact while writing the FBG. Both techniques are to produce a spatially localized (<50$\mu m$) periodic modulation of the core $N_r$ with a specific Bragg wavelength, $\Lambda_B$ and sufficient number of modulation periods to provide spectral selectivity, with depth of modulation determining $r_m(\lambda_o)$.

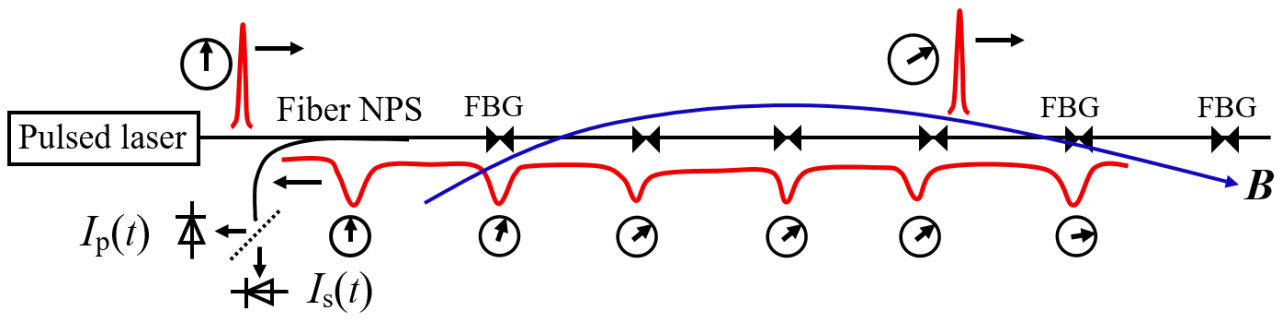

**Fig. 1** Schematic of an FOPP. A laser outputs vertically polarized pulses and a sequence of reflected pulses of energy $r_m E_o$ with spatially(temporally) varying SOP is generated by the sequence of FBGs. The backscatter is routed away from the laser using a non-polarizing fiber splitter. Malus' law determines $\alpha(s)$=$atan$($\sqrt{I_p/I_s}$)($t$). The PBS is oriented 45° to vertical for greatest sensitivity. The signal BW(1/$\tau_{det}$) would typically not resolve but smooth the intensity modulations.

One writing procedure: the Uniform BTOF, $r_m$=R, $m$=l,…,$N_p$, the return power, $P^1(t)$, is then exponentially decaying as shown in **Fig. 2**. Another writing procedure, the Flat BTOF, $P^1(t)$=$P^1(0)$, $r_m$ increases with $s_m$. This is shown in **Fig. 3**. For the Flat BTOF, the FBGs would be written with the return energy monitored in real time using a CW source. If 9 FBGs have been written, then the writing of the 10th FBG will be completed when the return signal rises by 1/9 of the previous return power. Then the fiber is translated to the 11th position and the procedure is repeated with 1/10 of return power. This writing procedure would compensate for two attenuation mechanisms: $\sigma_\lambda$[1/$m$] due to absorption at $\lambda$ and attenuation from earlier reflections as shown in **Fig. 2**.

The LIDAR reflected energy, d$E^1(m)$ from the $m^{th}$ FBG with reflection coefficient $r_m$ is given by Eq (3). The transmission coefficient, $t_m$=(1–$r_m$),

$$dE^1(m) = E_o r_m \prod_{i=0}^{m-1} t_i^2 \, , m = 1 \cdots N_p \, ; t_0 = 1 \qquad (3)$$

Reflected energy up to $m$, $E^1(m)$ is given by Eq (4),

$$E^1(m) = \sum_{j=1}^{m} dE^1(j) \qquad (4)$$

The instantaneous power is $P^1(m)$=d$E^1(m)/\Delta t$. The 0th order transmission coefficient, $T^0(m)$ for a fiber with $m$ FBGs is given by Eq (5),

$$T^0(m) = \prod_{i=1}^{m} t_i \, , \quad m = 1, \cdots, N_p \qquad (5)$$

A Uniform BTOF has a reflection series $r_m$=R, $t_m$=T=1–R. For a Uniform BTOF, the formulas simplify to d$E^1(m)$=$RE_o T^{(2m)}$, $E^1(m)$=$RE_o(1-T^{(2m+2)})/(1-T^{(2)})$ and $T^0(m)$=$T^{(m)}$. These distributions are shown in **Fig. 2** for a fiber with R=0.25%, $\Delta s$=5$cm$ and a 100nJ pulse energy.

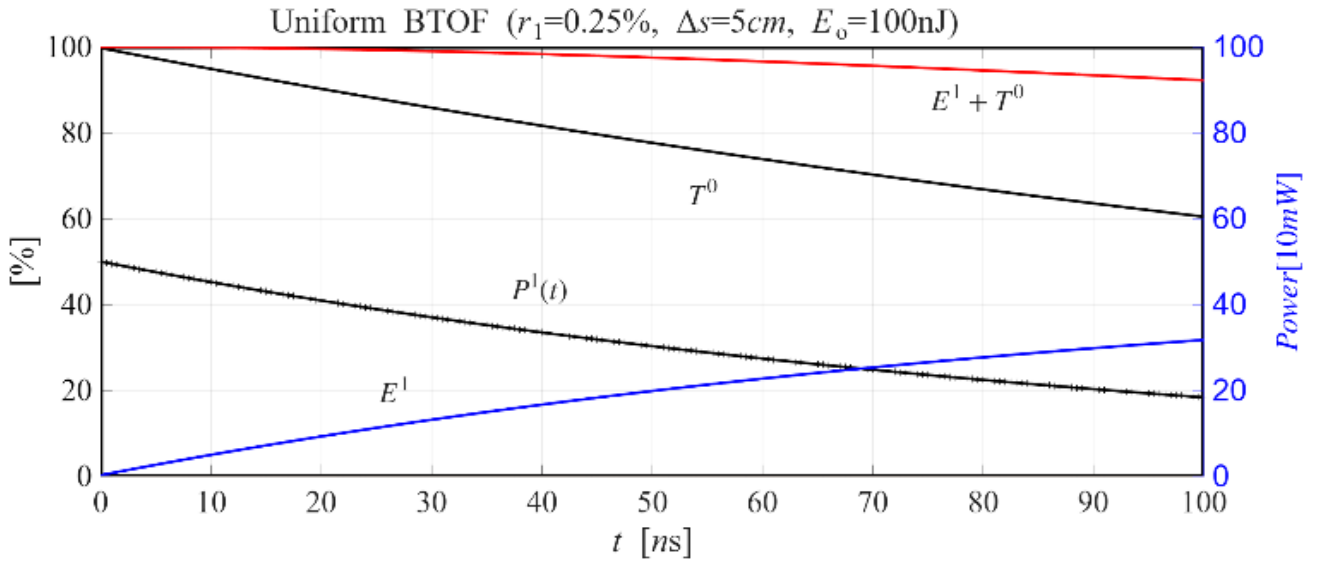

**Fig. 2** Uniform BTOF with $r$=0.25%, $\Delta s$ = 5$cm$, $\Delta t$=0.5$ns$, $N_p$=201, $E_o$=100$n$J. $P^1(t)$ is a digital exponentially decaying signal. The accumulated LIDAR return is $E^1(m)$. When added to the transmitted pulse energy $T^0(m)$, $E^1(m)$+ $E_o T^0(m)$<$E_o$.

A BTOF fiber becomes real when index is given a physical position, $s_m$=($m$–1)$\Delta s$. with $s$=$ct/2N_r$ and $\Delta t$=2$N_r \Delta s/c$. For $E_o$=100$n$J, $P^1(1)$=500$m$W, $\Delta s$=5$cm$ and $\Delta t$=0.5$ns$. $E^1(m)$ is exponentially decaying: $E_o exp(-2\sigma_s \cdot (m-1)\Delta s)$ with attenuation coefficient, $\sigma_s$= R/$\Delta s$(=0.05$m^{-1}$).

A Uniform BTOF has a dynamic range problem that can be corrected by tailoring the reflection series to produce a flat return. A Flat BTOF has an increasing $r_m$ with distance $s_m$ which keeps $P^1(t)$, shown in **Fig. 3**, with $r_1$=0.25%. The Flat BTOF has a fail point where $r_{N\text{p}}$ exceeds 10% to keep $\text{d}E^1(N_\text{p})=r_1E_\text{o}$. The algorithm for generating a Flat reflection sequence is given in **APPENDIX B.**

The unaccounted-for energy, $W(m)$ at the end of the Flat BTOF fiber is very large as reflection coefficients have increased to nearly 50%. $W(m)$ must be bracketed if the BTOF is to be optimized. One can decrease $r_1$ or the FBG density until $W(N_\text{p})<0.05\%$ say, but $P^1(t)$ and *SNR* are also reduced, perhaps unnecessarily.

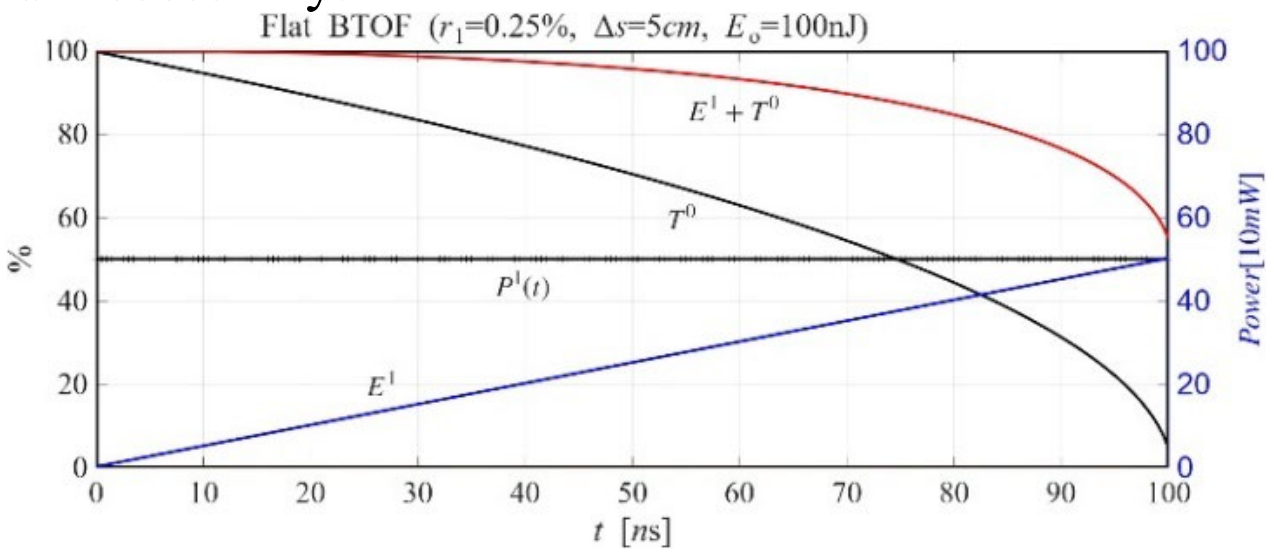

**Fig. 3** The first order accumulated reflected energy, $E^1$ and transmitted $E_\text{o}\text{T}^0$ in % for a Flat BTOF is shown. The unique sequence $[r_m]$ starts with $r_1$=0.25% with $r_\text{n}$ approaching 50% for the last FBG exhausting the pulse energy. $W=E_\text{o}-(E^1+E_\text{o}\text{T}^0)$ is large at the end of the fiber.

$W(m)$ is due to multi-paths: energy exiting the entrance of the fiber with 3, 5, … reflections $E^{3,5,\cdots}(m)$, contributing to the LIDAR signal and transmitted energy with 2, 4,… reflections given by $E_\text{o}\text{T}^{2,4,\cdots}(m)$ exiting the end of the fiber. The number of multipaths is strongly $m$ dependent with $W(m)$ peaking at $N_\text{p}$. $W(m)$ is innocuous at the beginning of the fiber.

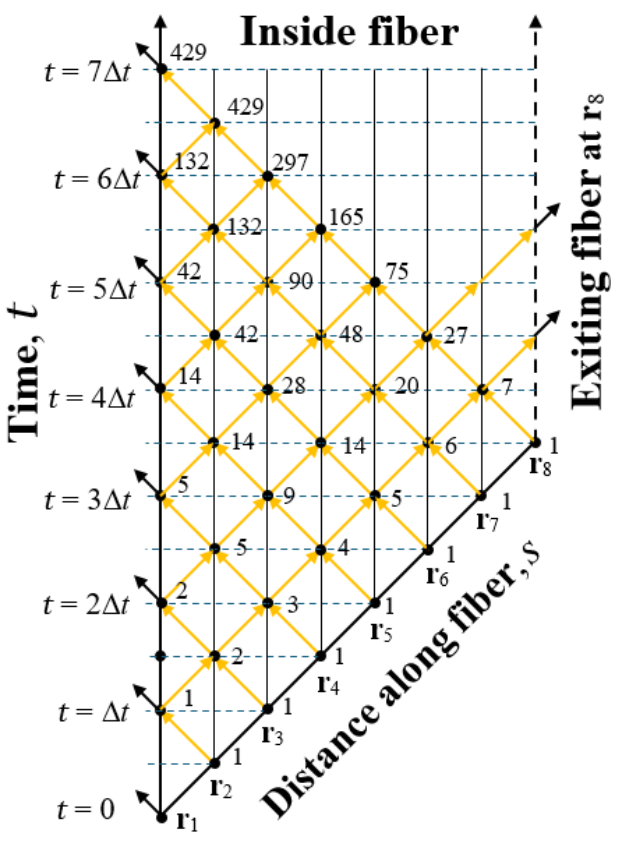

**Fig. 4** *Space–time* diagram of pulse trajectories in an 8 FBG BTOF. The # of paths contributing energy at the fiber entrance at times, $t=m\Delta t$ from all combinations of reflections is the Catalan series, $C_m^{2m}/(m+1)$, [1,2,5,14, 42,132,429,…], $C_7^{14}/8=429$. The prompt LIDAR signal are paths with only one reflection, $[r_m]$, $m$=1,…,8. The other pulses eventually exit the fiber at $r_8$ or decay to zero inside the fiber.

A *space–time* diagram of trajectories in a BTOF is shown in **Fig. 4**. The back-reflected energy at each discrete time, $t=(m-1)\Delta t$ can be computed in principle but the # of paths grows rapidly with time or $m$.

In **Figs 2** and **3**, $E_\text{o}\neq E^1(m)+E_\text{o}\text{T}^0(m)$ except for $m$=1. The fiber is assumed lossless, $W(m)$ contains the higher order paths, $E^n(m)$, $n$>1, $n$ odd and $E_\text{o}\text{T}^n(m)$, $n$>0, $n$ even. The number of trajectories exiting the entrance of the fiber at time $m\Delta t$ increases dramatically as $C_m^{2m}/(m+1)$. For $m$=174, there are 3.5x$10^{100}$ or 3.5 googel paths.

Without knowing the contamination from higher order reflections and what can be ignored as transmission, the measurement *SNR* along the fiber is in question. It was realized[4] that BTOFs can only be trusted over the first $N_\text{max}$ FBGs. By reducing $r$, $N_\text{max}$, and the useable length of the fiber increases. This paper provides the criteria for $N_\text{max}$ on a given BTOF design.

## III 3rd ORDER REFLECTED ENERGY: d$E^3(m)$

The total number of 3rd, 5th and higher order paths exiting the fiber entrance at $m\Delta t$ is generated using the algorithm given in **APPENDIX A.** This algorithm was derived by the author[4] but not pursued. Its significance as a generator of the *Narayana* numbers was not realized. The *Narayana triangle* has a diagonal sequence, [1, 3, 6, 10, 15, 21, 28], $m(m+1)/2$, $m$=l,...,7, the number of 3rd order paths exiting the fiber entrance at $(m+1)\Delta t$. Row-wise, the $n^\text{th}$ row of the Narayana triangle enumerate paths exiting at $(m+1)\Delta t$, which sum to Catalan sequence $C_m^{2m}/(m+1)$. Taking $m$=5: [1, 10, 20, 10, 1] there are 1–1st, 10–3rd, 20–5th, 10–7th and 1–9th order reflections exiting at time $6\Delta t$, $C_5^{10}/(6)=42$. *Generalized* Narayana numbers are also generated with no known relevance.

The 3rd order trajectories contain a $[r_ir_jr_k]$ product and $t_l^2$ factors as needed. An algorithm for summing all 3rd order reflected energies exiting at $(m-1)\Delta t$, $\text{d}E^3(m)$ is developed in **APPENDIX B** along with $\text{T}^{2,4}(m)$, the transmission coefficient of paths with $[r_ir_j]$ and $[r_ir_jr_lr_m]$ products, exiting the end of a fiber with $m$ FBGs.

The results for a Uniform BTOF are shown in **Fig. 5**. The useful length of fiber with *SNR*>50:1 is 3$m$ long, 60 FBGs. Using Tb doped glass, $\Delta\alpha=\text{V}_{\text{Tb},532nm}\Delta s_5=440°/\text{T}$. $B_\parallel<1\text{kG}$ to keep $\Delta\alpha<90°$. For silica fiber, $\Delta\alpha=\text{V}_{\text{Si},532nm}\Delta s_5=22°/\text{T}$. $P^1(0)=500m\text{W}$ with $E_\text{o}=100n\text{J}$.

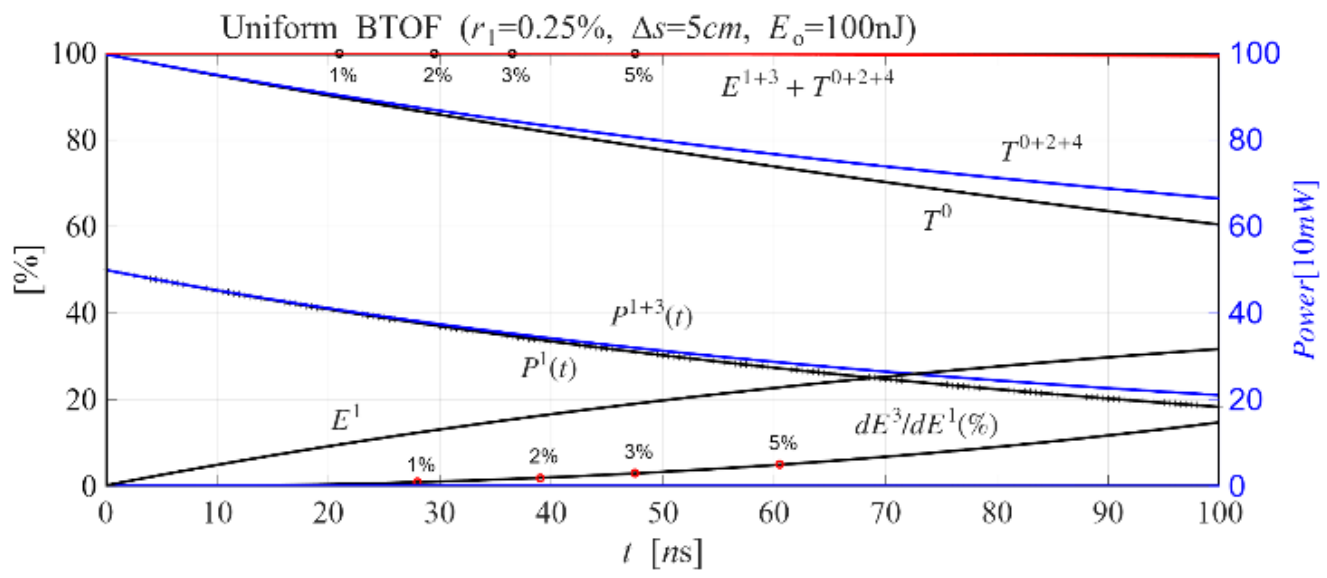

**Fig. 5** Dual traces $P^1$, $P^{1+3}(t)$ are shown for a Uniform BTOF with R=0.25%. The transmitted and reflected energies: $E_o T^0$, $E_o T^{0+2+4}$, $E^1$, $E^{1+3}$, $E_o T^{0+2+4}+E^{1+3}$ are plotted as %'s of $E_o$. *SNR* markers are placed at 1, 2, 3, 5% on $dE^3/dE^1$ and $(dE^3+dW)/dE^1$ where $W=E_o-(E^{1+3}+E_o T^{0+2+4})$. A 3*m* length with 60 FBGs satisfies *SNR*>50:1(upper % markers).

The results for a Flat BTOF are shown in **Fig. 6**. $N_{max}$ for the fiber with *SNR*>100:1 is ~40 with $L_f$= 2*m*.

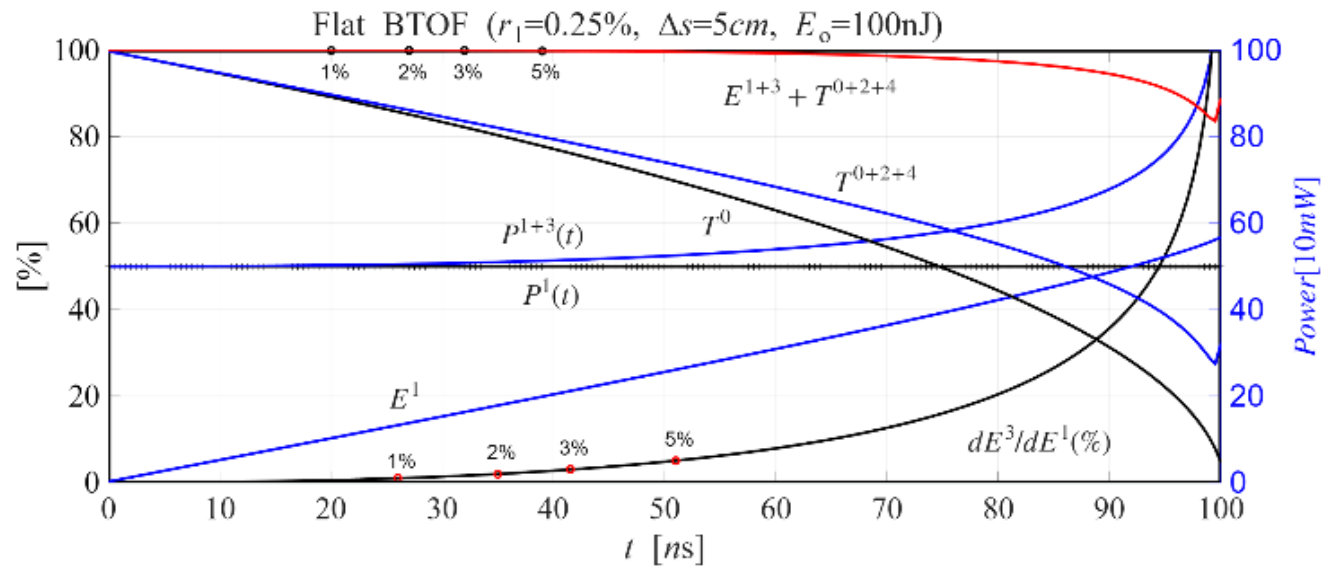

**Fig. 6** Flat BTOF with $r_1$=0.25%. Traces $P^1$, $P^{1+3}(t)$, $E^{1+3}+E_o T^{0+2+4}$, $E_o T^0$, $E_o T^{0+2+4}$ and $E^1$ as % of $E_o$. The *SNR* markers at 1, 2, 3 and 5% are given along the fiber for $dE^3/dE^1$ and for $(dW+dE^3)/dE^1$. $E^{1+3}+E_o T^{0+2+4}(N_p)$ is noticeably <100% of $E_o$.

**Fig. 6** shows the problem with Flat BTOFs. The usable range for a reflection series starting at 0.25% is about 30% of the fiber length with *SNR*<50:1. The errors contributing to the $dE^1$ are bracketed (upper markers %), nearly the same as $dE^3$ alone. The LIDAR return power is 500*m*W for $E_o$=100nJ with FBGs every 5*cm*.

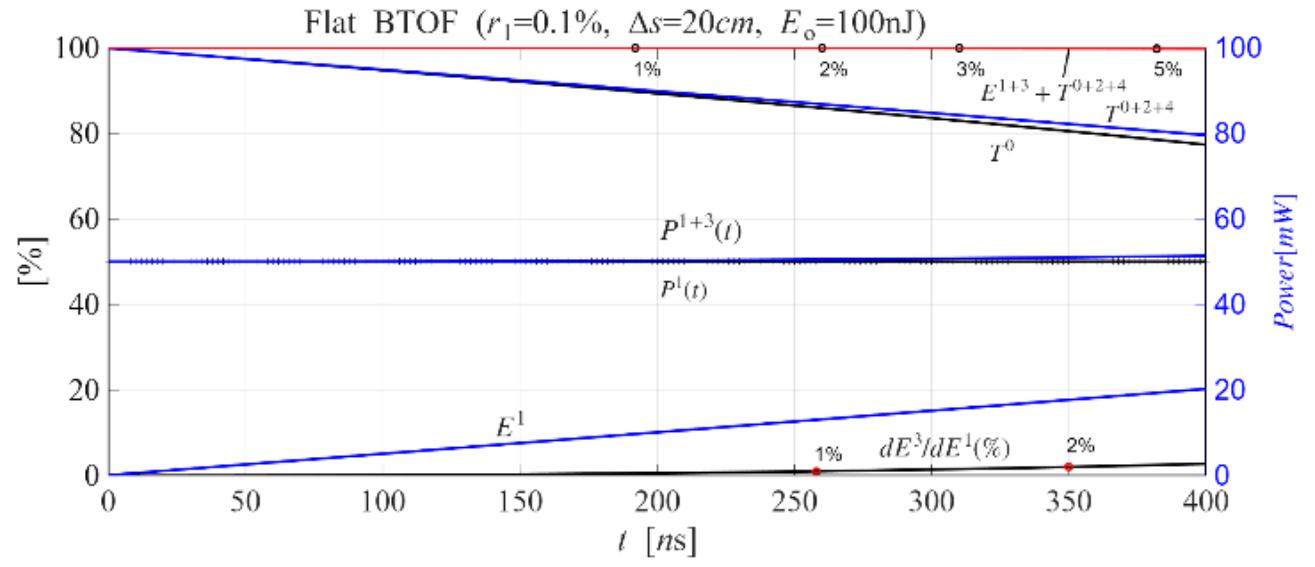

**Fig. 7** Flat return BTOF with $r_1$=0.1%. Traces of $P^1$, $P^{1+3}(t)$, $(E^{1+3}+E_o T^{0+2+4})$, $E_o T^0$, $E_o T^{0+2+4}$ and $E^1$ in % of $E_o$ are shown. The $dE^3/dE^1$ trace is marked at 1, 2% and $(dE-d(E^1+T^{0+2+4}))/dE^1$ is marked at 1, 2, 3 and 5%. The BTOF has SNR >50:1 for 25*m*. $P^1$= 50*m*W for $E_o$=100*n*J.

**Fig. 7** shows that Flat BTOF are practical at lower reflection, $r_1$=0.1%. The first reflection was reduced 2.5*x* to 0.1% and the FBG density was reduced, $\gamma$=4, $\Delta s$=20*cm*, resulting in a reduction of $P^1$ to 50*m*W.

## IV *SNR*, FBG DENSITY AND *B* FIELD RESOLUTION

The *spatial* field resolution $\delta B_{\|}/B_{\|}(s)$ along the fiber is lowered (better resolution) leading to higher return power if $\Delta s$ is reduced but then $\Delta\alpha$ is lower. The instrumental resolution of a well-made optical polarimeter is assumed to be ~0.1°. The fundamental noise level of detection is set by statistical shot noise with $N_{ph}$ in integration time $\tau_{det}$, $SNR_\gamma=\sqrt{QE}\sqrt{N_{ph}}$, detector QE is nearly 1 and taken to be 1. The detector's noise floor $\sim\sqrt{BW}(\sim\sqrt{1/\tau_{det}})$ can be limiting as LIDAR requires a high BW.

*Measurement SNR of BTOF*: If $r_m$ is reduced to the level that the first Born scattering approximation applies, $E^3(s)\approx0$, $dE^3/dE^1(m)$ is negligible then there is essentially no difference between a Uniform and Flat BTOF, the fiber length can be long but the return power is weak (strong relative to Rayleigh!). Taking $E_o$=1$\mu$J, the number of detected photons is $N^1_{532nm}=r\cdot2.7\cdot10^{12}E_{o,1}$ with $P^1(t)=rE_{o,1}/\Delta s_{10}$[kW]. $P^1$=10*m*W for $r=10^{-5}$. The $SNR_{532nm}\sim\sqrt{(N^1_{532nm}E_{o,1})}$:1 $=1.6\cdot10^6\sqrt{(rE_{o,1})}$ or $5\cdot10^3$:1 for $r=10^{-5}$. LIDAR requires a BW~$(1/\Delta s_{10})$[GHz]. A practical photodiode detector NEP, $P_n$, can be as low as 1fW/√Hz·√BW, even at GHz BWs with a $R_{det}=e\lambda_o/hc$=0.4A/W. $P_{n,det}$=0.03*n*W·$\sqrt{BW_{GHz}}$ for a $SNR_{det}=P^1/P_n$= $3\cdot10^{13}rE_{o,1}/\Delta s_{10}^{1/2}$:1 or $3\cdot10^8$:1 for $r=10^{-5}$. Measurement *SNR* is photon limited. A $r=10^{-7}$ achieves an *SNR*~600:1 which matches $SNR_{pol}$ with 0.1° resolution. In practice, weak FBG reflections of 0.1% or less are technologically demanding.

If the magnetic field $B_{\|}=\Delta\alpha/(2V_\lambda\Delta s)$ produces a rotation of several degrees (~10°) then $B_{\|}$ can be determined to ~1% accuracy. For silica fiber, $\Delta\alpha_{532nm}$=4.4°@1kG, $\Delta s$=10*cm*, resolving $B_{\|}$ to $\delta B_{\|}$=20G. Tb fiber is 20*x* more sensitive. The *local space–time* resolution is 1*ns*-10*cm* with rep rates of 5 MHz.

## IV FOPP INSTRUMENT DESIGN CONSIDERATIONS

*Waveform distortions from multi-pathing*: Placing a BTOF in a long solenoid with uniform field, $B_o$, all multi-paths exiting at $t=m\Delta t$ contribute identical Faraday rotations, $\alpha(m)$ as the relation between path length and exit time is fixed and by *non*–reciprocity, the Faraday effect depends only on path length not path direction. The multi-pathing can only be decerned in $\alpha(m)$ $(=\alpha^{1+3}(m))$ if the $B_{\|}(s)$ is nonuniform. To this end, a $B_{||}(s)$ with a pure sinusoidal variation is applied to a Uniform BTOF in **Fig. 5**, the results are shown in **Fig 8**. To accomplish this, the $dE^3(m)$ algorithm is decomposed into elemental paths, $dE^3([i,j,k],m)$ to provide an energy weighted line integrated Faraday rotation, $dE^3([i,j,k],m)$ $\alpha^3([i,j,k],m)$ for that specific path. The energy averaged Faraday rotation for a measurement is then given Eq (6), and the algorithm measured for a given $B_{||}(s)$ along the BTOF, $\alpha(m)$ is detailed in **APPENDIX B**,

$$\alpha(m) = \frac{\left(2\alpha^1 dE^1 + \sum_{[ijk],m} \alpha^3 dE^3\right)(m)}{(dE^1 + dE^3)(m)} \qquad (6)$$

It is found that towards the end of the Uniform BTOF, $B_m(s)$ is progressively attenuated relative to $B_{||}(s)$. Phase distortion(shift in zero-crossings) is small. This would seem to imply that a BTOF can measure complex $B_{||}(s)$ profiles and, once Fourier decomposed, be corrected, allowing the entire fiber to be used with confidence.

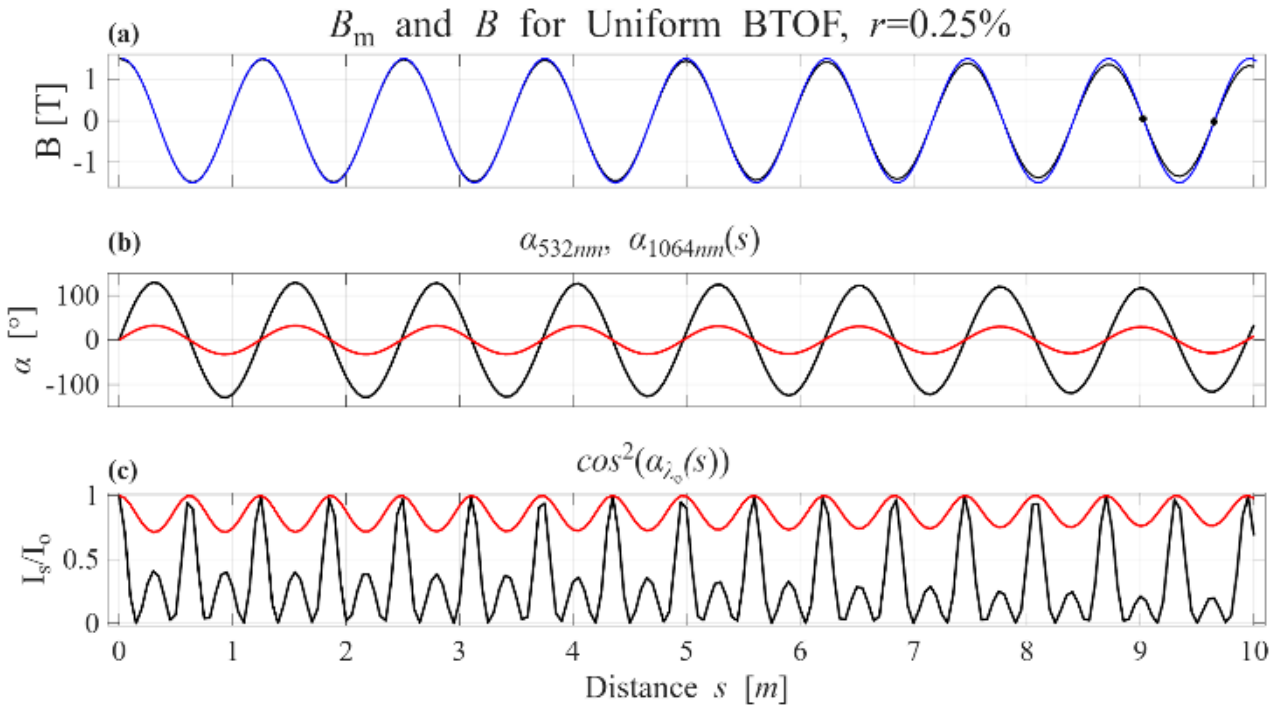

**Fig. 8** Since a complex $B_{||}(s)$ can be Fourier decomposed, a logical test field is a pure sine wave (8 cycles). In 8**(a)**, $B_m(s)$ with $s$ shifted by $-\Delta s/2$ is shown to be attenuated by ~9% relative to $B_{||}(s)$. The markers are placed where traces coincide, phase distortion is low. The measured Faraday rotation, $\alpha(s)$, 8**(b)** is shown for $\lambda_o$=[532,1064]$nm$, 8**(c)** plots $cos^2\alpha(s)=I_s(\lambda_o,s)/I_o$. $I_p(\lambda_o,s)/I_o=1-cos^2\alpha(s)$ has no new information.

This demonstrates that the errors introduced by 3rd order and higher multi–pathing are not stochastic but systematic. The true $B_{||}(s)$ can be retrieved from the measured $B_m(s)$ if the BTOF fiber is well characterized.

*Determination of $N_{max}$* : The energy in the multi-paths of a BTOF can be quantified by appealing to the fiber's exit. The transmitted energy, $E_o T(t)$ is ideally a single pulse exiting at $t=N_p\Delta t/2+t_p$ with energy $E_o T^0(N_p)\sim E_o$ and Faraday rotation, $\alpha_T=V_\lambda \langle B\rangle_{Lf}\cdot L_f$, highlighting that an FOPP serves as a FOCS measuring $I_p(t_p)$ as shown in **Fig. 10** on a tokamak. Multipathing, if evident, will generate a distributed transmitted signal in time. The transmitted power $P_T(t)$ will be consequences of the BTOF design and will indicate where along the fiber the multi-pathing becomes an issue.

*Polarimeter considerations*: The sensor section of the FOPP is to be straight, or nearly straight SMF to minimize introducing stress birefringence due to bending the fiber. This introduces a linear birefringence, $\chi(s)$ which should be reduced well below $\alpha(s)$. Lo-Bi and spun fibers are used to minimize the intrinsic residual stress in the fiber from pulling the fiber. Larger devices have higher $B_{||}$ and gentler bends which reduces the binding stress. FBGs can be written on specialty fibers. PMFs are used to transmit the light pulse to the BTOF and collect return distributed signal from the BTOF after the non-polarizing splitter.

The FOPP instrument has a **no**–$B$ signal as a baseline for subsequent scans with coils and plasma energized. The author is of the mind that the pulsed source should be a *dithered* polarized source, with dithering between two orthogonal LP states, say ±45°. Every pulsed source can be so configured, in principle. **Fig. 9** shows an arrangement where two pulses are generated for each pulse by a polarization based optical delay. The delay, $\tau_D$ separates the pulses by at least $\Delta t_p > 2N_r L_f/c$ to guarantee that only one back-reflected signal is measured at a time.

A single detector polarimeter has advantages over a conventional two detector polarimeter. In a two-detector system, the detectors must be '*balanced*' in responsivity and dark current with an insensitivity to mechanical drift, (beam displacement on detector's active area). Deviations are hard to characterize. A single detector is automatically balanced. Angles are not but the difference of intensities normalized by the sum of intensities, *difference–over–sum*. For this method, $\alpha(t)$ is given in Eq (5) for ±45°LP states and $\alpha(t)$<45°,

$$2\alpha(t) = \left(I_{45°} - I_{-45°}(\tau_D)\right)/\left(I_{45°} + I_{-45°}(\tau_D)\right) \qquad (5)$$

where the intensities are measured at times $t$ and $t-\tau_D$. The *local* field $B_{||}$ is given by Eq (6),

$$B_{||}^*(s) = \frac{1}{4V_\lambda}\Delta\left(\frac{I_{45°} - I_{-45°}(\tau_D)}{I_{45°} + I_{-45°}(\tau_D)}\right)/\Delta s \qquad (6)$$

$B^*_{||}(s)$ is the field profile produced by using a smooth differentiable curve to fit the measured $\alpha(s)$ or $I_{s,p}(s)$ profiles and differentiating. A nonlinear formula is used for larger $\alpha(s)$. Replicating this scheme with a detector on the second analyzer output would fully utilize the laser pulse energy and increase the $SNR$ by √2.

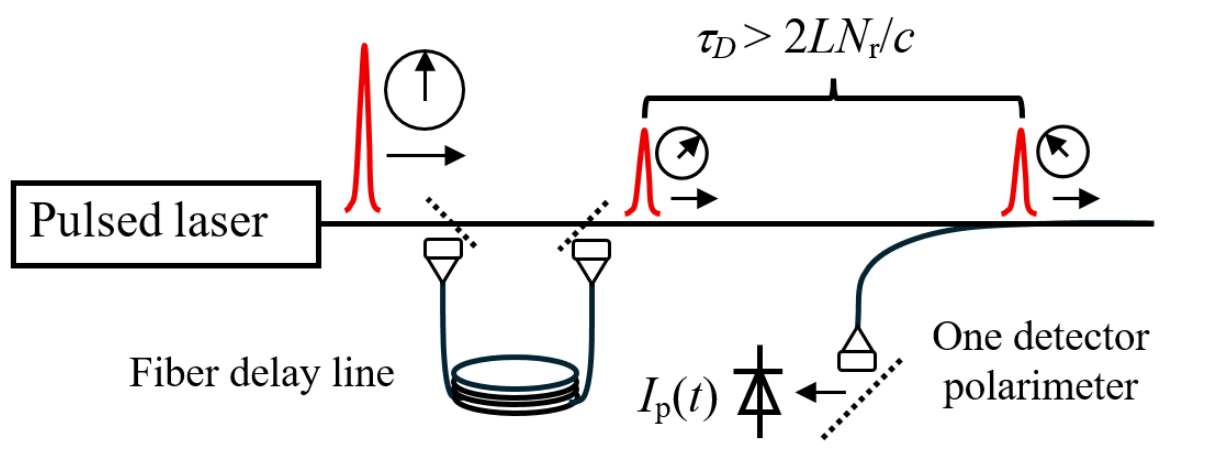

**Fig. 9** Illustration of a single detector polarimeter from a *dithered* orthogonally polarized two pulse source using a polarization based fiber delay. The PBS oriented 45° splits off, delays (by $\tau_D$) and reintroduces half of the injected pulse with little loss. *Distributed* $\alpha(s)$ is measured directly by *difference–over–sum* intensity method using one detector.

The electronics and digital signal processing FPGA carries out a *difference–over–sum* scheme in *real–time*. The parameter $B_{\|}^*(s)$ is directly measured which facilitates prompt *real–time* feedback for plasma control.

The appealing aspect of a one detector technique is its transparency. The polarimeter is being aligned and used at both orthogonal states reducing instrumental errors. For a **no**–$B$ scan, the difference must be zero as the intensities are equal for a well aligned polarimeter and LP states symmetric about the analyzer's axis. Two scans are used for a *distributed* $B_{\|}(s)$ measurement, the two pulses are almost coincident in time. Laser rep rates can be 1–5MHz, $\tau_D$=0.1$\mu$s for a 10$m$ fiber.

## V Practical Applications of FOPP in MFE

Fiber optic current sensor(FOCS) technology has been envisioned on MHD plasmas for decades. The plasma current, $I_p(t)$, is the most direct indicator of the plasma discharge's success but electrical magnetic sensor measurements of $I_p$ can walk off due to integrator drift over time or saturate for transient bursts. A FOCS doen't have this problem. The main problem with FOCS is both intrinsic and bending induced birefringence fading out the Faraday rotation but this has been solved with low birefringent optical fibers[8]. Fibers with beat length, $L_B$ greater than 150$m$ can accurately measure $I_p$ on ITER, fibers with $L_B$ >300$m$ are available. Since then, spun fibers have been characterized and shown to be suitable[9]. FOCS is being developed for ITER[10] and has been implemented on the JET tokamak[11,12]. There has been a decade long development of sorting out key issues of magnetic sensors on a burning plasma where a high neutron fluence is detrimental to electrical probes but also optical fibers, radiation induced absorption(RIA). Fortunately, damaged fibers can be replaced in situ when necessary if re-entrant metal tubes are used as housings. As an example, of a FOPP applied to ITER, a poloidal FOPP is shown in **Fig. 10**, Taking the loop circumference to be 18$m$ and an average $B_{pol}$ of 1T, $\Delta\alpha_{Si,532nm}$=23°$\Delta_5$@1T with $N_p$=360. One can reduce $N_p$ to 180 if $\lambda_o$=1064$nm$ and $\Delta s$=10$cm$, then $\Delta\alpha_{Si,1\mu m}$=12°$\Delta s_{10}$. The back reflected emission duration, $\Delta t_f$=180$ns$ so a laser rep rate of 1/$\Delta t_f$~5 MHz is appropriate. If a *dithered* source is used as shown, the rep rate is reduced to 2.5 MHz. The field resolution can be 1% of the ambient field. The $\delta$<$B_{pol}(s)$> can be reduced during post–processing in two ways: 1) by boxcar averaging $N$ scans, increasing the *SNR* by $\sqrt{N}$ or 2) using a running average over $N$ FBGs improving the local sensitivity by $\sqrt{N}$ while still covering the poloidal or structure spatial mode numbers of interest.

The FOPP measures $B_{pol}(\theta)$ at discrete $\theta_m$, $m$=1,…,$N_p$, providing input to Grad-Shafranov equilibrium codes[13] with $I_p(t_p)\propto\alpha(L_f,t_p)$ or $\alpha_T(t_p)$ but also *local* or segmented field or current measurements. A direct electrical analogue to an FOPP, that seems to be finding favor, are the discrete, 'partial' or segmented Rogowski coils, as implemented on the West tokamak[14] using 330 vessel and 44 divertor Rogowski coils. The plasma current is then $I_p(t)=\sum_{i=1}^{Np}\alpha_i B_{tang,i}(t)$ with $B_{tang}$ a partial Rogowski coil measurements. The KSTAR device[15] uses $B_{dot}$ probes to do the same, $\mu_o I_p(t)=\sum_{i=1}^{N} B_{pol,i}(t)\Delta s_i$ for startup and plasma breakdown where high BW is appreciated. This provides a direct comparison of the two concepts: the BW of Rogowski coils is ~kHz; integrators are required if not *self*-integrating(low BW); cabling is demanding for such large arrays and Mirnov arrays are not replaced which require high BW. FOPP provides even higher probe densities. The *distributed* FOPP measurements provide boundary shape and shape parameters: Shafranov shift $\Delta_{GS}$ ($m_\theta$ =1) and vertical/horizontal displacements. The X–point topology is also a target for FOPP and partial Rogowski coils. For equilibrium measurements the FOPP scans can be boxcar averaged to reduce the noise floor and increase *local* SNR level. The up-front costs of FOPP more than compensate for their advantages.

For fluctuations the FOPP excels as a high BW~MHz *local, distributed* diagnostic. A poloidal FOPP acts like a Mirnov array detecting fluctuations, $\delta B_{pol}(\theta_m,t_p)$ to very high high poloidal mode number $m_\theta$ . A BTOF wound helically around the plasma as illustrated in **Fig. 10**, provides fluctuation measurements of both $m_\theta$ and $n_\varphi$ mode numbers in specific ratios, $m_\theta$:$n_\varphi(s)$. In **Fig. 10** a curious helical FOPP is wound that nulls out the helical equilibrium field to detect weak magnetic fluctuations. If a substantial ***B*** field is sensed, then the local pitch of ***B*** relative to the FOPP is noted.

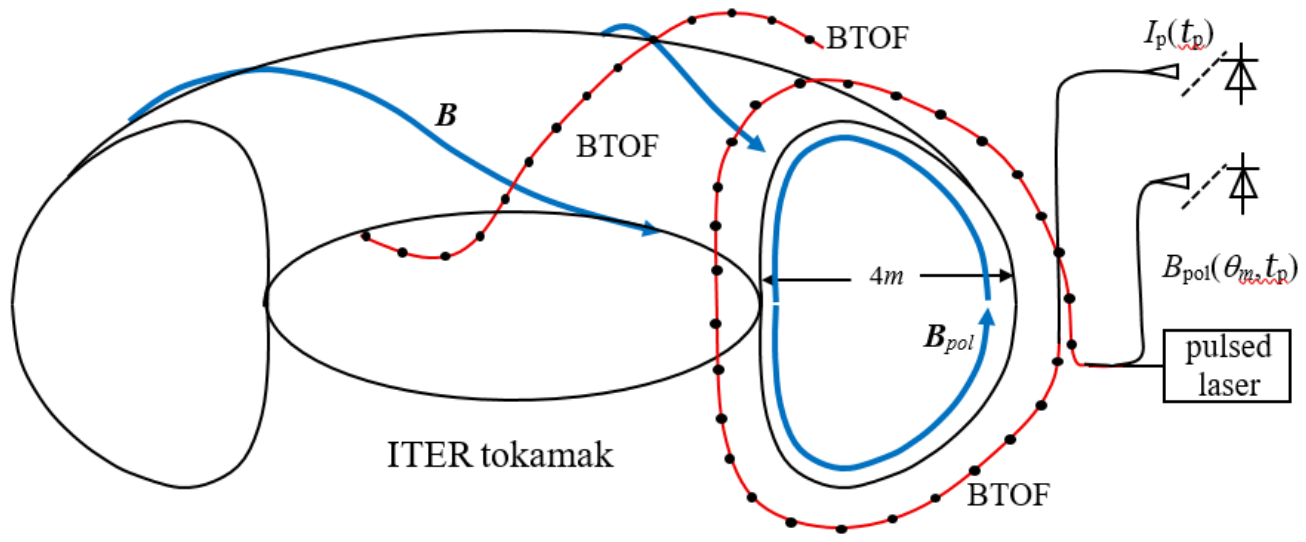

**Fig. 10** Illustration of a FOPP on ITER tokamak. The poloidal field is ~1T. With $B_{tor}$=5.5T on axis, $R_o$=6.2*m*, $a_o$=2*m*, the vessel. The plasma current, $I_p$ is 15MA.

These fluctuations appear external to the current source (plasma) in vacuum but diminish with distance from the plasma(16). The localization of the fluctuations using distributed field sensing is important for detecting problems during startup, rotating magnetic island modes, motions in a particular direction and can detect instabilities where the discharge needs to be terminated with a BW of 3-5MHz. Also, disruptions can saturate electrical magnetic sensors with integrators, fiber sensing is robust to this, with better machine protection.

These scenarios apply to all MHD equilibria.

## VI Conclusion

The author has revisited the subject of BTOFs and fiber optic pulsed polarimetry to satisfy a curiosity about the sequences discovered in [4]. The author's algorithm has been shown to produce the *Catalan* series for total paths exiting the fiber with odd # of reflections and the partial series for a specific odd # are found to be *Narayana triangular* numbers. The algorithm also generated the *generalized Narayana* numbers. Maybe the first 'practical' application that uses these numbers.

The specific BTOF designs in **Figs. 5** and **6** are useful as they are adaptable. For instance, a Uniform BTOF with reflection of 0.125%, half of R can have a doubled FBG density, $\Delta s$=0.25*cm* over 3*m*, $N_{max}$=120 with SNR> 50:1 and this design can be adapted with a scaling, $\gamma$ to any fiber length. Only uniformly spaced BTOFs were considered but the density can be changed along the fiber as well as the reflectivity to make the anticipated field measurement more leveled. The LIDAR discrete distribution in time will be more complicated as the multi-paths will generate a finer comb of exit times.

The energy, $W(m)$ contaminating the LIDAR response has been bracketed by showing that its largest component, the 5th order multi-path energy, is negligible for a given design. Uniformly spaced BTOFs can now be designed and used with confidence. A single detector *dithered* source Fiber Optic Pulsed Polarimeter design has been presented.

FOPP has been shown to build on, not replace, present FOCS technology while adding *distributed* sensing. The measurement bandwidths are high for demanding transient phenomena but easily averaged to a post processed low bandwidth. The performance for *real–time* feedback is best for a LIDAR technique, which is useful for resolving plasma disruptions, plasma startup and important for device protection. The *B* field is *space–time* resolved to 10*cm–ns* at high rep rates. *Robust* measurements of vacuum *B* are practical using the FOPP technique with BTOFs.

This paper also gives the author the opportunity to catch up on fiber optic magnetic sensor development in the MFE program. It is logical that several programs are using segmented Rogowski sensors to provide *local* current measurements but surprising in such high numbers and densities and that POTDR using Rayleigh scattering has been revived(17) which was introduced as a *distributed* field fiber sensor for *km* long transmission lines, even then, with boxcar averaging many scans. It is hoped that *fiber optic pulsed polarimetry* using *backscatter-tailored optical fibers* will find a place on present and future devices. Ref [4] was not written to suggest that FOPP is limited to MHEDLP devices but also covers the large spectrum of laboratory plasma and transient high intensity *B* field research.

## APPENDIX A: Multipaths and the Narayana Triangular Numbers

The pattern of multi–pathing reflections is most suggestive if displayed as in **Fig. Al** where the paths are seen to increase in recursive patterns with each successive level off depth in BTOF. Old patterns(red) increase as {0, 1, 3, 5, 7, …}, similar patterns(blue) arise later {0, 0, 0, 1, 3, 5, 7} shifted by 2 as do more interwoven multipaths. The following algorithm for **H** and **V** sequences was suggested by these patterns.

For $n=N_p-1$, **H** and **V** are $n\cdot(n+1)\cdot 2n$ matrices initialized as follows:

**H**(l:*n*, 1:*n*+l, 1:2*n*) = **V**(l:*n*, 1:*n*+l, 1:2*n*) = 0;
**V**(l:*n*, 1, 1) = l; **V**(l, 2, 3) = l; **H**(l:*n*, 1, 2) = l;

The *algorithm*:

```
for j = 2: n;    for k = 2: j;
        V(j, k, :) = sr(H(   j,  k–1, :)) + V( j–1,    k, :);
        H(j, k, :) = sr(V( j–1,     k, :)) + H(    j, k–1, :);
                  end;
  H(j, j+1, :) = H( j,  j, :);  V(j, j+1, :) = sr(H( j,  j, :));
end;
```

where **sr**($[a_1, \ldots, a_{2n}]$)=$[0, a_1, \ldots, a_{2n-1}]$ is a shift register. The outputs to $n$ = 8 are,

**H**(2, 2, :)=[0, 1, 0, 1, 0, 0, ... ] 1-1st and 1-3rd order.
**H**(3, 3, :)=[0, 1, 0, 3, 0, 1, 0, 0, ... ] 1-1st, 3-3rd and 1-5th order.
**H**(4, 4, :)=[0, 1, 0, 6, 0, 6, 0, 1, 0, 0, ... ]
**H**(5, 5, :)=[0, 1, 0, 10, 0, 20, 0, 10, 0, 1, 0, 0, ... ]
**H**(6, 6, :)=[0, 1, 0, 15, 0, 50, 0, 50, 0, 15, 0, 1, 0, 0, ... ]
**H**(7, 7, :)=[0, 1, 0, 21, 0, 105, 0, 175, 0, 105, 0, 21, 0, 1, 0, 0]
**H**(8, 8, :)=[0, 1, 0, 28, 0, 196, 0, 490, 0, 490, 0, 196, 0, 28, 0, 1]

Elements of **H**($j, j$, :) are given by $C_{k-1}^{j-1} \cdot C_{k-1}^{j}/k$, $k$=1:$j$, $C_{k-1}^{7} \cdot C_{k-1}^{8}/k$ = [1, 28, 196, 490, 490, 196, 28, 1] for $j$=8.

**H**($j, j$, :) gives the number of multi-paths of odd order for $n$ FBG reflectors. How each multi-path contributes to a given LIDAR output requires further investigation.

**V**(2, 2, :)=[0, 0, 2, 0, … ]
**V**(3, 3, :)=[0, 0, 2, 0, 3, 0, ... ]
**V**(4, 4 ,:)=[0, 0, 2, 0, 8, 0, 4, 0, .. ]
**V**(5, 5, :)=[0, 0, 2, 0, 15, 0, 20, 0, 5, 0, ... ]
**V**(6, 6, :)=[0, 0, 2, 0, 24, 0, 60, 0, 40, 0, 6, 0, ... ]
**V**(7, 7, :)=[0, 0, 2, 0, 35, 0, 140, 0, 175, 0, 70, 0, 7, 0, ]
**V**(8, 8, :)=[0, 0, 2, 0, 48, 0, 280, 0, 560, 0, 420, 0, 112, 0, 8, 0]

The non-zero components of **V**($j, j$, :) are given by $2C_{k}^{j+1} \cdot C_{k-1}^{j-2}/(j+1)$; [2, 48, 280, 560, 420, 112, 8] for $j$=8, $k$=1:7. The **H** and **V** matrices are orthogonal partitions(sum to) of the Catalan series: $C_{j}^{2j}/(j+1)$=[1, 2, 5, 14, 42, 132, 429, ... ] row-wise.

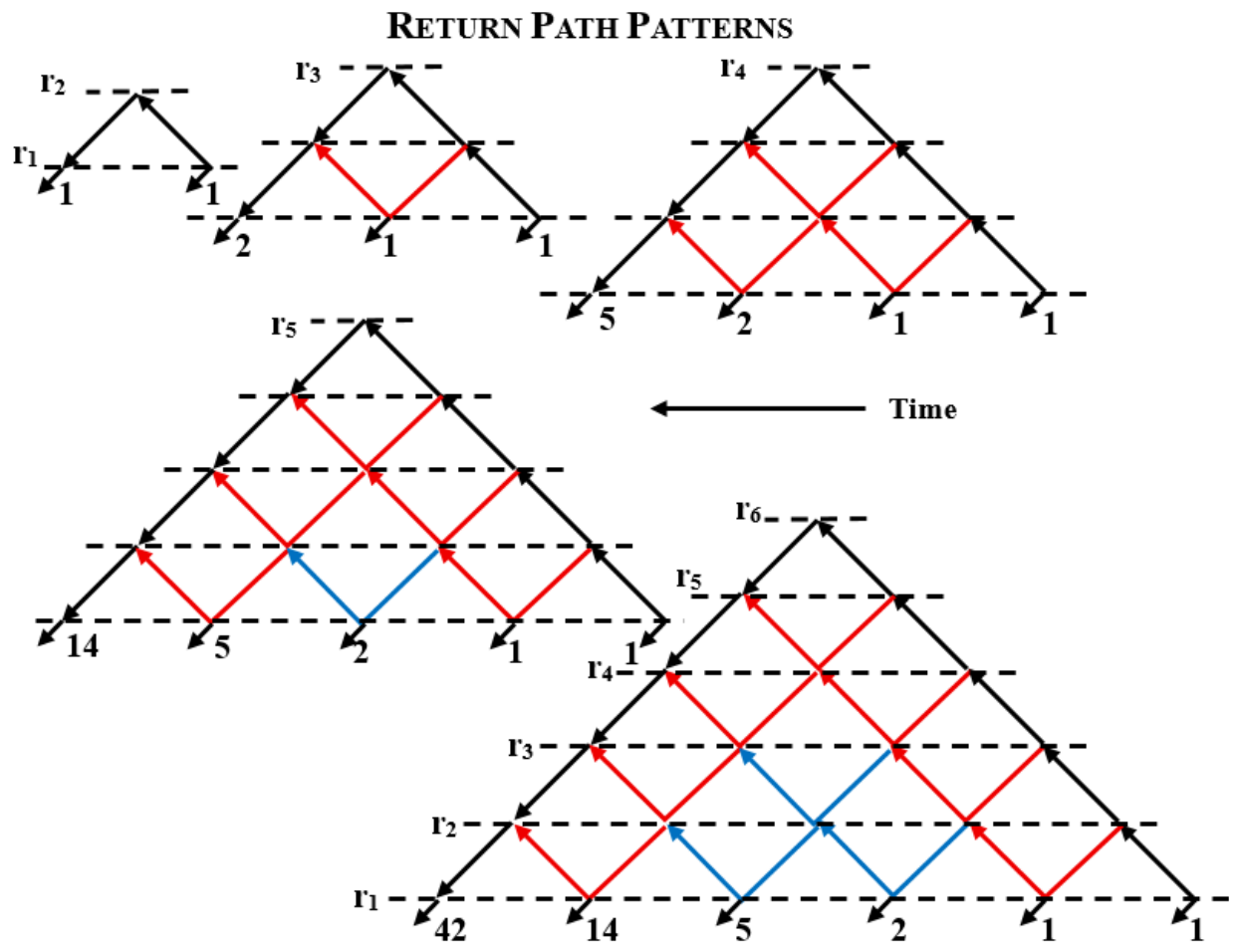

**Fig. A1** Trajectory patterns that develop the *Narayana* numbers(# of reflections). Patterns repeat while gaining in number(red) suggesting a recursive formula. The patterns increase vertically and horizontally by 1 with each step.

The triangular patterns of **H**($j, j$, :) and **V**($j, j$, :) are displayed in **Fig. A2**.

The 3rd order reflection paths contributing to d$E^1(m)$ are numbered by the 2nd diagonal series, [1, 3, 6, 10, 15] or $m(m+1)/2$ for $m$=1, 2, 3, 4, 5.

The 5th order reflection paths contribute to d$E^5(m)$ are enumerated by the 3rd diagonal series, [1, 6, 20, 50, 105,…] or $m^2(m^2-1)/12$ for $m$= 2, 3, 4, 5, 6, … . The pattern is:

$1^2 = 1$
$1^2+(1^2+2^2) = 6$
$1^2+(1^2+2^2)+(1^2+2^2+3^2) = 20$
$1^2+(1^2+2^2)+(1^2+2^2+3^2)+(1^2+2^2+3^2+4^2) = 50$
$1^2+(1^2+2^2)+(1^2+2^2+3^2)+(1^2+2^2+3^2+4^2)+(1^2+2^2+3^2+4^2+5^2) = 105$

Related are the 2nd and 4th order paths that transmit energy out the end of the fiber. $T^2(m)$ and $T^4(m)$. for a BTOF with $m$ FBGs.

1
1 2 1
1 2 3 3 1
1 2 6 8 6 4 1
1 2 10 15 20 20 10 5 1
1 2 15 24 50 60 50 40 15 6 1
1 2 21 35 105 140 175 175 105 70 21 7 1
1 2 28 48 196 280 490 560 490 420 196 112 28 8 1

**Fig. A2** The *Narayana* numbers are given by **H**($m,m$,:) and **V**($m,m$,:) is interlaced and underlined. The *generalized Narayana* numbers, **V**($n, n$, :) have yet to find a physical interpretation.

## APPENDIX B: CALCULATIONS OF d$E^3$, $T^{2,4}$ and $\alpha(m)$

*Algorithm* for generating a Flat reflection series:

*Initialize*: $N_p = 201$; $r(1) = 0.0025$; $t(1)=1-r(1)$;
T2(1)=$t$(1)^2; d$E^1$(1)=$E_0 \cdot r(1)$;
**for** $m$= 1, $N_p$–1;
  $t(m+1)$ = 1–$r(m)$;
  T2($m$+1) = T2($m$) · $t$($m$+1)^2;
  $r(m+1)$ = $r$(1)/T2($m$+1);
  d$E^1(m+1)$ = $E_0$T2($m$+1) · r($m$+1);
**end;**

$N_p$ is chosen where $r_{N_p}$~0.5, if $r_1$>0.0025 then $N_p$<201.

*Algorithm* for generating d$E^3$:

*Rules* for d$E^3(m)$ [$r_i, r_j r_l$] exiting at $t=m\Delta t=m\ 2\Delta s N_r/c$.

1) $pt_1$ to $r_i$ and $pt_1$ to $r_l$ involve $(i-1)$ and $(l-1)t^2$ factors
2) $j < i, l$
3) $2(i+l-j-1)=2m$, $m$=2:$N_p$

The d$E^3$ and $T^2$ multi-paths are illustrated in **Fig. B1** for a 4 FBG BTOF.

Referring to **Fig. B1(a)**, the LIDAR signal d$E^1$(1:4) at $t$=[1, 2, 3, 4]$\Delta t$ are given by Eq (3), in **B1(a)**, d$E^3$(3:4)= $E_o[t_1^2 r_2 r_1 r_2,\ (t_1t_2)^2\{r_2r_1r_3+r_3r_1r_2+r_3r_2r_3\}]$ at $t$=[2,3]$\Delta t$; $T^2(4)=t_4t_1[r_2r_1t_2t_3+t_2r_3r_2t_3+t_2t_3r_4r_3t_3,\ t_2r_3r_1t_2t_3+t_2t_3r_4t_3r_2t_3,\ t_2t_3r_4t_3t_2r_1t_2t_3]$ at $t$=[5,7,9]$\Delta t$/2. To be complete, $T^2(2)=[t_1t_2r_2r_1]$, $T^2(3)=[t_1t_3\{r_2r_1t_2+t_2r_3r_2+t_2r_3t_2r_1t_2]$ and $T^0(1:4)=[t_1,\ t_1t_2,\ t_1t_2t_3,\ t_1t_2t_3t_4]$.

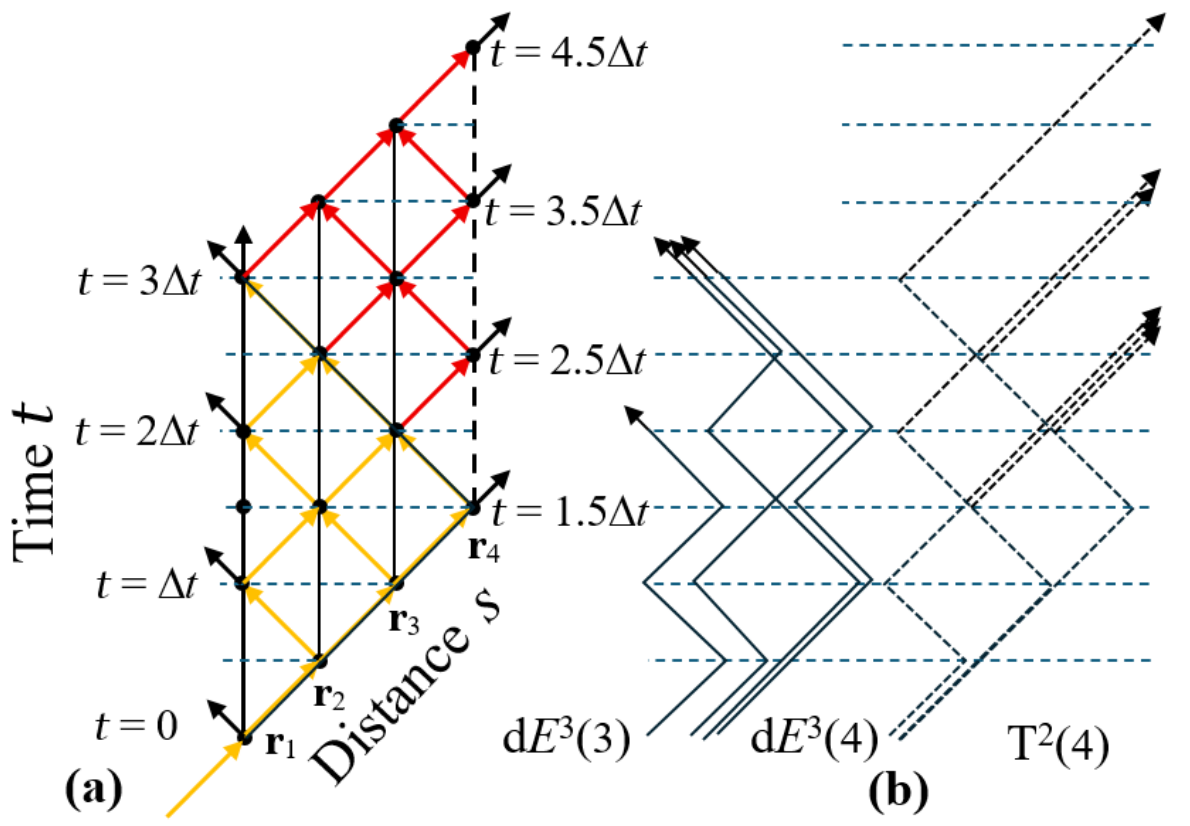

**Fig. B1, B1**(a), a *space–time* diagram of the laser pulse for a 4 FBG BTOF. In **B1**(b), $dE^3(3)$, $dE^3(4)$ total 1 and 3 paths. In **B1**(b), $T^2(4)$ has 6=4(4–1)/2 paths(dotted), the first reflection is followed by 1, 2, and 3 reflections in order to exit the end of the fiber. $T^2(4)$ is *distributed* from 5/2–9/2Δ*t*.

Multipathing in an 8 FBG fiber is shown in **Fig. B2**. where the {[4,3,7], [7,3,4]} and [7,6,7] paths are shown, reflection positions $r_i$, $r_j$ and $r_k$ denoted by [*i*,*j*,*k*].

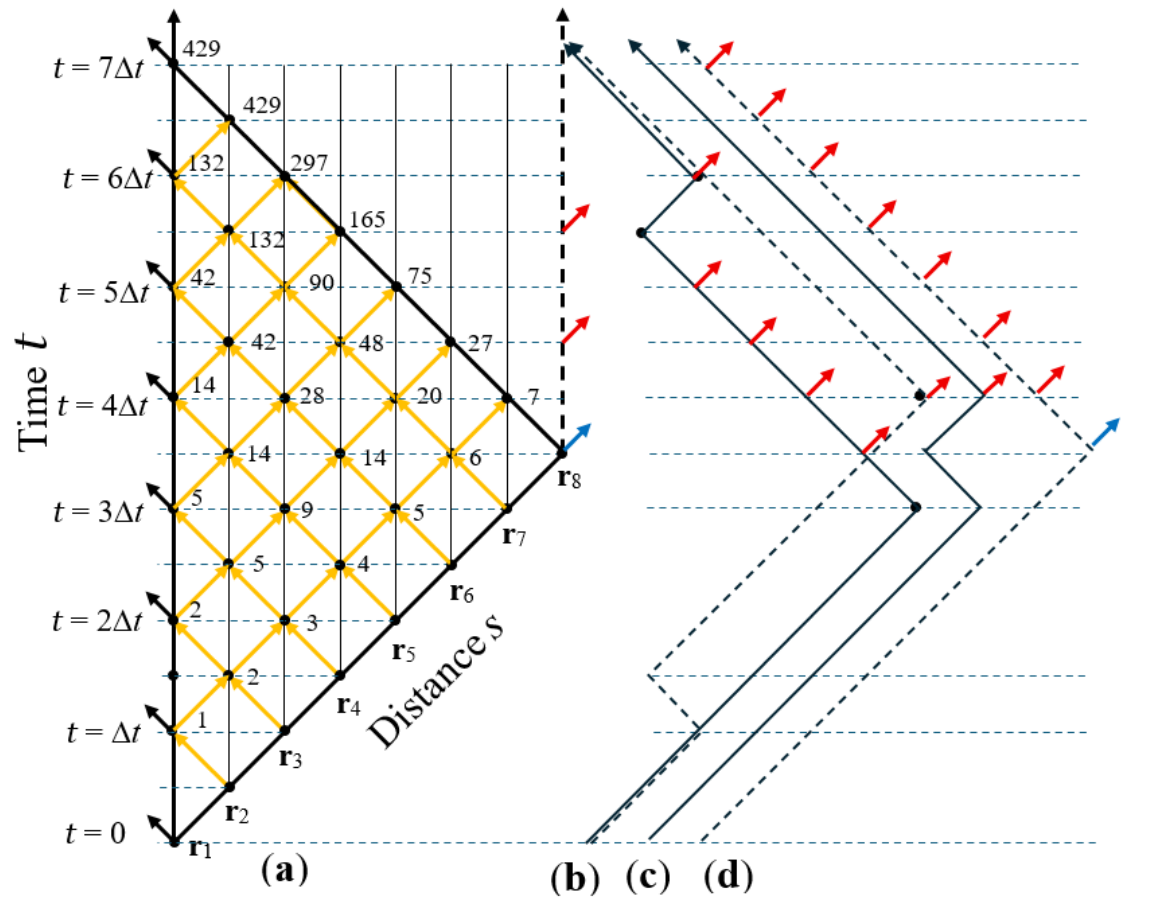

**Fig B3** (**a**) shows a *space–time* diagram of an 8 FBG BTOF, 429 paths up to 13th order contributing to d*E*(8), with transmitted paths exiting fiber. **(b)** shows $dE^3(8)$ paths that are pair symmetric $[r_3r_2r_7]$ and $[r_7r_2r_3]$ and a self-symmetric path $[r_7r_6r_7]$ of which there are 3. The $T^2(8)$ paths are **not** identical on pair symmetric paths. The $dE^1(8)$ single path is shown (**d**). $T^2(8)$ paths are started by turning back to the end of the fiber after one reflection as shown by short arrows. The $T^0(8)$ path is shown on (**a**, **d**) with no reflections.

T1(*m*) and T2(*m*) are defined in Eq (B1),

$$\mathrm{T1}(m)=\prod_{i=1}^{m} t_i \text{ and } \mathrm{T2}(m)=\mathrm{T1}(m)^2, m=1,\cdots,N_p \quad \text{(B1)}$$

For $[r_ir_jr_k]$ path, Let matrix **rr**(*j*, *k*) denote a reflection at $r_j$ and then propagating to and reflecting at $r_k$, Eq (B2),

$$\mathbf{rr}(j,k)=r_jr_k\frac{\mathrm{T2}(k-1)}{\mathrm{T2}(j)}, j=1:N_p-1, k=j+1:N_p \quad \text{(B2)}$$

illustrated by $\mathbf{rr}(9,30)=r_9r_{30}\, t_{10}^2\cdot\ldots\cdot t_{29}^2$.

The **trace** operator is used to sum the diagonal entries of the $\mathbf{rr}(1{:}1{+}j,k{:}k{+}j)_{(1+j),(1+j)}$ square matrix by Eq (B3),

$$\mathbf{trace}(\mathbf{rr}(1:1+j,k:k+j))=\sum_{i=1,1+j}\mathbf{rr}(i:i+k-1) \quad \text{(B3)}$$

The 3rd order reflection coefficients, $dE^3(1,\ldots,8)$ are:

$dE^3(1)=dE^3(2)=0$;
$dE^3(3)=E_or_2(\mathbf{rr}(1,2)\qquad)\cdot\mathrm{T2}(1)$
$dE^3(4)=E_or_3(\mathbf{rr}(2,3)+2\cdot\mathbf{trace}(\mathbf{rr}(1{:}1,2{:}2))\qquad)\cdot\mathrm{T2}(2)$
$dE^3(5)=E_or_4(\mathbf{rr}(3,4)+2\cdot\mathbf{trace}(\mathbf{rr}(1{:}2,2{:}3))\qquad)\cdot\mathrm{T2}(3)$
$\qquad+r_3(\mathbf{rr}(1,3)\qquad)\cdot\mathrm{T2}(2)$
$dE^3(6)=E_or_5(\mathbf{rr}(4,5)+2\cdot\mathbf{trace}(\mathbf{rr}(1{:}3,2{:}4))\qquad)\cdot\mathrm{T2}(4)$
$\qquad+r_4(\mathbf{rr}(2,4)+2\cdot\mathbf{trace}(\mathbf{rr}(1{:}1,3{:}3))\qquad)\cdot\mathrm{T2}(3)$
$dE^3(7)=E_or_6(\mathbf{rr}(5,6)+2\cdot\mathbf{trace}(\mathbf{rr}(1{:}4,2{:}5))\qquad)\cdot\mathrm{T2}(5)$
$\qquad+r_5(\mathbf{rr}(3,5)+2\cdot\mathbf{trace}(\mathbf{rr}(1{:}2,3{:}4))\qquad)\cdot\mathrm{T2}(4)$
$\qquad+r_4(\mathbf{rr}(1,4)\qquad)\cdot\mathrm{T2}(3)$
$dE^3(8)=E_or_7(\mathbf{rr}(6,7)+2\cdot\mathbf{trace}(\mathbf{rr}(1{:}5,2{:}6))\qquad)\cdot\mathrm{T2}(6)$
$\qquad+r_6(\mathbf{rr}(4,6)+2\cdot\mathbf{trace}(\mathbf{rr}(1{:}3,3{:}5))\qquad)\cdot\mathrm{T2}(5)$
$\qquad+r_5(\mathbf{rr}(2,5)+2\cdot\mathbf{trace}(\mathbf{rr}(1{:}1,4{:}4))\qquad)\cdot\mathrm{T2}(4)$

The pair symmetric paths have identical coefficients. The $dE^3$ coefficients have even/odd patterns as follows:

For *k* even, *m*=[(*k*+2)/2]=*k*/2+1:
$dE^3(k)/E_o=r_{k-1}(\mathbf{rr}(k-2,k-1)+2\cdot\mathbf{trace}(\mathbf{rr}(1{:}k-3,2{:}k-2)))\cdot\mathrm{T2}(k-2)$
$\qquad+r_{k-2}\cdot\mathbf{rr}(k-4,k-2)+2\cdot\mathbf{trace}(\mathbf{rr}(1{:}k-5,3{:}k-3)))\cdot\mathrm{T2}(k-3)$
…
$\qquad+r_m\cdot\ \mathbf{rr}(2,m)+2\cdot\mathbf{trace}(\mathbf{rr}(1{:}1,m-1{:}m-1))\qquad)\cdot\mathrm{T2}(m-1)$

For *k* odd, *m*=[(*k*+2)/2]=(*k*–1)/2+1:
$dE^3(k)/E_o=r_{k-1}\cdot\mathbf{rr}(k-2,k-1)+2\cdot\mathbf{trace}(\mathbf{rr}(1{:}k-3,2{:}k-2)))\cdot\mathrm{T2}(k-2)$
$\qquad+r_{k-2}\cdot\mathbf{rr}(k-4,k-2)+2\cdot\mathbf{trace}(\mathbf{rr}(1{:}k-5,3{:}k-3))\ )\cdot\mathrm{T2}(k-3)$
…
$\qquad+r_{m+1}\cdot\mathbf{rr}(2,m+1)+2\cdot\mathbf{trace}(\mathbf{rr}(1{:}1,m-1{:}m))\qquad)\cdot\mathrm{T2}(m)$
$\qquad+r_m\ \cdot\mathbf{rr}(1,m)\qquad)\cdot\mathrm{T2}(m-1)$

This is illustrated by $dE^3(8)$ and $dE^3(9)$:

$dE^3(8)/E_o=\ r_7\cdot(\mathbf{rr}(6,7)\ +2\cdot\mathbf{trace}(\mathbf{rr}(1{:}5,2{:}6))\ )\cdot\mathrm{T2}(6)$
$\qquad+r_6\cdot(\ \mathbf{rr}(4,6)\ +2\cdot\mathbf{trace}(\mathbf{rr}(1{:}3,3{:}5))\ )\cdot\mathrm{T2}(5)$
$\qquad+r_5\cdot(\ \mathbf{rr}(2,5)\ +2\cdot\mathbf{trace}(\mathbf{rr}(1{:}1,4{:}4))\ )\cdot\mathrm{T2}(4)$
$dE^3(9)/E_o=\ r_8\cdot(\mathbf{rr}(7,8)\ \ +2\cdot\mathbf{trace}(\mathbf{rr}(1{:}6,2{:}7))\ )\cdot\mathrm{T2}(7)$
$\qquad+r_7\cdot(\ \mathbf{rr}(5,7)\ \ +2\cdot\mathbf{trace}(\mathbf{rr}(1{:}4,3{:}6)))\cdot\mathrm{T2}(6)$
$\qquad+r_6\cdot(\ \mathbf{rr}(3,6)\ \ +2\cdot\mathbf{trace}(\mathbf{rr}(1{:}2,4{:}5)))\cdot\mathrm{T2}(5)$
$\qquad+r_5\cdot(\ \mathbf{rr}(1,5)\qquad)\cdot\mathrm{T2}(4)$

The accumulated 1st, 3rd order reflected energy sequences are given by Eq (B3),

$$E^{1,3}(m)=\sum_{i=1}^{m} dE^{1,3}(i) \quad \text{(B3)}$$

*Algorithm* for generating *α*(*m*):

Define $\alpha^1\mathrm{ij}(i,j)$, the Faraday rotation between indices *i*, and *j* anywhere in the fiber given the applied $B_{\|}(s)$,

**for** $m=1{:}N_p$; $d\alpha^1(m)=V_\lambda\Delta s\cdot(B_{\|}(m\Delta s)+B_{\|}(((m-1)\Delta s))/2$; **end**;
Initialize: $\alpha^1\mathrm{ij}(1{:}N_p,1{:}N_p)=0$;
**for** $i=1{:}N_p$; **for** $j=i{:}N_p$; $\alpha^1\mathrm{ij}(i,j)=\mathrm{sum}(d\alpha^1(i{:}j))$; **end; end;**

Find $dE^3(m)$ again using elemental paths:

Initialize: $dE^3$=zeros(1:$N_p$); $d^3alp^1$=zeros(1:$N_p$); $dE^3alp^1$=zeros(1:$N_p$);

**for** $m$=2:$N_p$; $dE^3(m) = 0$;

**for** $i$=1:$N_p$; **for** $j$=1:$N_p$; **for** $j$=1: min($i$, $j$)–1;

**if**($i$+$k$–$j$–1 == $m$);

$dE^3(m)=dE^3(m)+t1jk(1,i-1)\cdot r(i)\cdot t1ijk(j+1,i-1)\cdot t1jk(j+1,k-1)\cdot r(j)\cdot t1jk(1,k-1)\cdot r(k)$;

$d^3\alpha(m)=\alpha^1ij(1,i-1)+\alpha^1ij(j,i-1)+\alpha^1ij(j,k-1)+\alpha^1ij(1,k-1)$;

$dE^3\alpha^3(m)=dE^3\alpha^3(m) + dE^3(m)\cdot d^3\alpha(m)$;

**end**; %end on if

**end**; **end**; **end**;

**end**;

As a check, $dE^3(m)$ matches previous $dE^3(m)$.

$\alpha(m)=(dE^1(m)\cdot 2\alpha^1ij(1,m-1)+dE^3\alpha^3(m))./(dE^1(m)+dE^3(m))$.

Measured $B$: $B_m(m+1)=(1/2V_\lambda)(\alpha(m+1)-\alpha(m))/\Delta s$. $B_m(1)$ set equal to $B_m(2)$, illustrated in **Fig .8**.

This approach in deriving $dE^3(m)$ does not lead to the *Narayana triangular* numbers. Straight forward algorithms miss the hidden beauty uncovered with labor!

TRANSMISSION COEFFICIENTS: $T^{0,2,4}(m)$

Fiber transmission is dominated by the straight through path, no reflections, $T^0(m)=[t_1\cdot\ldots\cdot t_m]$. $E_oT^0(N_p)$ appears at $t=L_fN_r/c$. The transmission, $\alpha_{T0}$ is a line integrated quantity, $\alpha_{T0}=V_\lambda[B_{||}]_{Lf}$, exiting at $T_p+L_fN_r/c$.

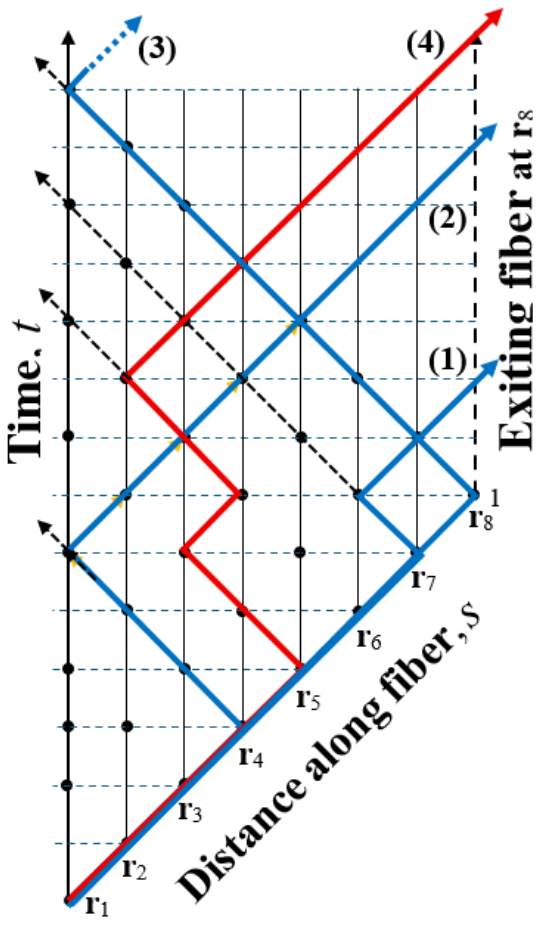

**Fig. B4** The transmission trajectories of an 8 FBG BTOF. Four paths are shown exiting the fiber at distributed times. Three of the paths(1–3) belong to $T^2(8)$ and one(4) to $T^4(8)$. Note: the transmission paths are the same as paths belonging to $dE^1$ and $dE^3$ but with an added reflection on the leg that exits towards the fiber entrance(dotted lines). The $T^2(8)$ distribution over time is from the shortest time is $9\Delta s/2$ to $7\Delta s+7\Delta s/2$.

Three paths for 2nd order $T^2(8)$ are shown in **Fig. B4** as well as one path contributing to transmitted pulses exit at different times $T^4(8)$. Transmitted pulse energy, $E_oT^2(m)$ depend on linear products of the $t_i$ and there is hope that they contribute substantially to $W$. Interesting combinatoric formulas may result based on the observations in **Fig. B4** that $T^{2,4}(m)$ are related to the return $dE^{1,3}(m)$ but $T^2(m)$ and $T^4(m)$ will calculated using nested loops over all possibilities.

*Rules* for $T^2(m)$: reflections at $r_k$ first and then $r_j$

1) $k$=2:$m$
2) $j$=1:$k$–1
3) Pulse moves to $pt_m$ after $r_j$ and exits with coefficient $t_m$

Define $T1_{jk}$, linear products of the $t_j$, a square matrix:

*Initialize:* $T1_{jk}(j,k) = 1$, $j$ = 1:$N$p, $k$ =1:$N$p

**for** $m$ = 1:$N_p$; $T1_{jk}(m,m)= t(m)$; **end**;

**for** $k$ = 1, $N_p$; **for** $j$ = 1: $k$–1;

**if**($j$ == 1); $T1jk(j, k) = T1(k)$;

**else;** $T1_{jk}(j, k) = T1(k)/T1(j-1)$;

**end**; %end on if else

**end**; **end**;

*Algorithm* for $T^2(m)$

*Initialize*: $T^2(1)$=0;

**for** $m$ = 2, $N_p$; $T^2(m)$=0;

**for** $k$ = 2:$m$; **for** $j$ = 1: $k$–1;

$T^2(m)= T^2(m)+T1_{jk}(1, k-1)\cdot r(k)\cdot T1_{jk}(j+1, k-1)\cdot r(j)\cdot T1_{jk}(j+1, m)$;

**end**; **end**;

**end**; %end on $m$

**Fig. B3**, paths as a check:

**(1)** $j$=1, $k$=8, $T^2(8)=T^2(8)+T1_{jk}(1,6)\cdot r_7\cdot T1_{jk}(7,6)\cdot r_6\cdot T1_{jk}(7,8)$

**(2)** $j$=1, $k$=4, $T^2(8)=T^2(8)+T1_{jk}(1,3)\cdot r_4\cdot T1_{jk}(2,3)\cdot r_1\cdot T1_{jk}(2,8)$

**(3)** $j$=6, $k$=7, $T^2(8)=T^2(8)+T1_{jk}(1,7)\cdot r_8\cdot T1_{jk}(2,7)\cdot r_1\cdot T1_{jk}(2,8)$

The $T^2(m)$ have been extensively checked.

*Rules* for $T^4(m)$: reflections at $r_k$, $r_j$, $r_{kk}$ and $r_{jj}$

1) $k$=2:$m$
2) $j$=1: $k$–1
3) $kk$=$j$+1: $m$
4) $jj$=1: $kk$–1
5) Pulse moves to $pt_m$ after $r_{jj}$ and exits with coefficient $t_m$

*Algorithm* for $T^4(m)$

*Initialize:* $T^4(1{:}2)$=0;

**for** $m$ = 3, $N_p$; $T^4(m)$=0;

**for** $k$=2:$m$; **for** $j$=1:$k$–1; **for** $kk$=$j$+1:$m$; **for** $jj$=1:$kk$–1;

$T^4(m)=T^4(m)+T1_{jk}(1,k-1)\cdot r(k)\cdot T1_{jk}(j+1,k-1)\cdot r(j)\cdot T1_{jk}(j+1, kk-1)\cdot r(kk)\, T1_{jk}(jj+1, kk-1)\cdot r(jj)\, T1_{jk}(jj+1, m)$;

**end**; **end**; **end**; **end**;

**end**; % end on $m$

The energy that lies outside of $E^{1+3}$, $T^{0+2+4}$ is $W(m)=E_o-(E^{1+3}(m)+E_oT^{0+2+4}(m))$. $E^5(m)$ is the largest contribution to $W(m)$. The effects of $E^3(m)+W(m)$ on the measurement *SNR* is illustrated for Flat and Uniform BTOFs in **Figs 5** and **6**, where $N_{max}$ from *SNR* % markers due to $dE^3(m)+dW(m)$ is compared to $N_{max}$ from *SNR* % markers due to $dE^3(m)$ alone.

## **APPENDIX C** Notation, terminology and simple relations

***Physical Constants:***

| | |
|---|---|
| $c, h, e$ | speed of light $3\cdot10^8[m/\mathrm{s}]$, Planck's constant $6.64\cdot10^{-34}$[Js], electron charge 1.602e-19[C] |

***Acronyms:***

| | |
|---|---|
| LIDAR, SOP, Tb | LIght Detection And Ranging, state of polarization, terbium |
| SMF, PMF, LP | single mode fiber, polarization maintaining fiber, linearly polarized |
| FOPP, MHD, MFE | Fiber optic Pulsed Polarimeter, magneto-hydrodynamic, magnetic fusion energy, |
| FRC, RFP, ELM | Field reversed configuration, reverse field pinch, edge localized mode |
| DAQ, NIR, SRS, SBS | data acquisition system, , near infra-red, stimulated Raman, Brillouin scatter |
| DTL, FR, FOPP, CW | damage threshold limit, Faraday, fiber optic pulsed polarimetry, continuous wave |
| BTOF, POTDR | backscatter tailor optical fiber, polarization optical time domain reflectometry |
| FBG, UV, RMP | fiber Bragg grating, ultraviolet, resonant magnetic perturbations, |
| $SNR$, $SNR_{\gamma,det,pol}$ | measurement signal-to-noise ratio, $SNR$(due to photon, detector, polarimeter-noise) |
| FWHM, BW, BP, FPGA | full width at half max, FWHM bandwidth, bandpass, field-programmable gate array |
| NPBS(NPS), PBS, MHEDLP | (non-) polarizing beam splitter, Magnetized high energy density laboratory plasma |
| FOCS, NIR, Lo-Bi | fiber optic current sensor, near infrared, low birefringence optical fiber |

***Light pulse parameters:***

| | |
|---|---|
| $\lambda_o, f_o(=c/\lambda_o)\ (\psi,\chi)(s)$ | wavelength$[m]$, frequency[Hz], pulse SOP(polarization azimuth, ellipticity) along fiber$[rad]$ |
| $\tau_p$, $l_p(=c\,\tau_p)$, | duration$[s]$, length$[m]$ |
| $(\psi_o, \chi_o)$, $E_o$, $P_o(=E_o/\tau_p)$, $\boldsymbol{E}_p$ | initial SOP$[rad]$, energy[J], power[W], electric field[V/$m$] |

***BTOF Fiber Optic Pulsed Spectropolarimeter instrumental parameters and subject:***

| | |
|---|---|
| $t_p$, $\Delta t_f$, $\Delta s$, $\Delta t(=2\Delta s N_r/c)$ | pulse time, duration of backscatter$[s]$, FBG spacing(spatial resolution)$[m]$, spacing in time$[s]$ |
| $s(=ct/2N_r)$, $\hat{\mathbf{s}}$, $L_f$, $\sigma_s$ | distance along fiber$[m]$, fiber axis direction, fiber length$[m]$, exponential decay constant $[1/m]$ |
| $pt_m$, $r_m$, $t_m$ | point $m$, coefficient of reflection, transmission at $pt_m$ |
| $\Delta t_p$, $P^{1,3}(t)$ | Time delay between orthogonally polarized pulses$[s]$ , 1st and 3rd order signal powers[W] |
| $\tau_B$, $l_B$, $\tau_{det}$, | field dynamic time scale$[s]$ and length of fiber restricted by $\tau_B[m]$, detector response time$[s]$ |
| $\Delta_{GS}$, $m_\theta$, $n_\varphi$ | Grad Shafranov shift$[m]$, poloidal and toroidal mode numbers |
| $N_{ph}$, $\Lambda_B$ | # of photons in $\tau_{det}$, Bragg grating wavelength$[m]$ |
| $dE^{1,3}(m)$, $dW(m)$ | 1st and 3rd order reflected energies from $pt_m$ on BTOF[J], differential of $W(m)$[J] |
| $E^{1,3}(m)$, $\Delta s_{10}$, $E_{o,1}$ | Accumulation of $dE^{1,3}$ to $pt_m$ on BTOF[J], $\Delta s$ and $E_o$ in units of $10cm$ and $1\mu$J |
| $W(m)(=E_o-E_o T^{0+2+4}-E^{1+3})(m))$ | unaccounted for accumulated energy[J] composed of $E^{5+7+\cdots} + E_o T^{6+8+\cdots}(m)$[J] |
| $T^{0,2,4}(m)$,T | 0th , 2nd and 4th order transmission coefficients to $pt_m$ on BTOF, total transmission, $T^{0+2+\cdots}$ |
| $E^{1,3}(m)$, $dE^3/dE^1$, | Accumulated return energy to $m^{th}$ FBG[J], $SNR$ of BTOF using $dE^3$ as error |
| $N_{r,(si,Tb)}$ | index of refraction of fiber's glass (silica, Tb-doped glass) |
| $\tau_D$, $\gamma$, $L_f$, $N_p(=L_f/\Delta s+1)$ | delay between polarized pulses$[s]$, scaling factor for BTOF, length of BOTF$[m]$, # of FBGs |
| $\alpha_f(s)$, $\alpha(s)$, $\psi(s)$ | accumulated Faraday rotation at $s[rad]$, at detector$[rad]$, polarization azimuth in fiber$[rad]$ |
| $\boldsymbol{B}$, $B_{\|}(s)$, $I_{p,s}(t)$ | magnetic field[T], $\boldsymbol{B}$ field component in direction of $\hat{\mathbf{s}}$ along fiber[T], analyzed intensities[W] |
| $V_{\lambda(si,Tb)}$, | fiber's wavelength dependent Verdet constant for (silica, Tb) glass$[rad/\mathrm{T}-m]$, |
| $N^1{}_\lambda(m)$ | # collected photons with wavelength $\lambda$ from reflection $r_m$ |

***Detection parameters:***

| | |
|---|---|
| $P_{n,det}$, QE, $R_{det}$ | NEP or noise floor[W], quantum efficiency(assumed 1), responsivity[A/W] |